\documentclass[aps,twocolumn,superscriptaddress,preprintnumbers,showpacs, floatfix, nofootinbib]{revtex4-1}
\usepackage{amsfonts,amssymb,graphicx,psfrag,dsfont,slashed,bigints}
\usepackage{amsfonts,amssymb,graphicx,times,slashed, xcolor,bm}
\usepackage[colorlinks=true,citecolor=blue]{hyperref}
\usepackage[normalem]{ulem}
\newcommand{\de}{\partial}

\newcommand{\be}{\begin{equation}}
\newcommand{\ee}{\end{equation}}
\newcommand{\ba}{\begin{eqnarray}}
\newcommand{\ea}{\end{eqnarray}}

\begin{document}

\title{Dissipative processes at the acoustic horizon}

\author{Maria Luisa Chiofalo}
\affiliation{Dipartimento di Fisica, Universit\`a di Pisa,
Polo Fibonacci, Largo B. Pontecorvo 3}
\affiliation{INFN Sezione di Pisa,
Polo Fibonacci, Largo B. Pontecorvo 3, 56127 Pisa, Italy}

\author{Dario Grasso}
\affiliation{INFN Sezione di Pisa,
Polo Fibonacci, Largo B. Pontecorvo 3, 56127 Pisa, Italy}

\author{Massimo Mannarelli}
\affiliation{INFN, Laboratori Nazionali del Gran Sasso, Via G. Acitelli,
22, 67100 Assergi (AQ), Italy}
\affiliation{Gran Sasso Science Institute, Viale Francesco Crispi, 7, 67100 L'Aquila}

\author{Silvia Trabucco}
\affiliation{Gran Sasso Science Institute, Viale Francesco Crispi, 7, 67100 L'Aquila} 
\affiliation{INFN, Laboratori Nazionali del Gran Sasso, Via G. Acitelli,
22, 67100 Assergi (AQ), Italy}

\begin{abstract}
A transonic fluid flow  generates an acoustic hole that is the hydrodynamic  analogue of a gravitational black hole. Acoustic holes emit a detectable thermal radiation of phonons at a characteristic Hawking temperature. The crucial concept is that the spontaneous phonon emission  at the horizon  produces an irreversible heat increase  at the expenses of  the bulk fluid kinetic energy. We show that such process can be described in terms of \textit{effective} shear and bulk viscosities that are defined close to the horizon.
We analyze this quantum friction process by resorting to a general kinetic theory approach as well as by the specific  description of  phonon emission as a tunneling process. The celebrated Kovtun,  Son  and  Starinets (KSS) universal lower bound $\eta /s = 1 / 4 \pi $ of the shear viscosity coefficient to entropy density ratio, readily follows, and is extended to the longitudinal bulk viscosity at the horizon. We come to the same saturation of the KSS bound after considering the shear  viscosity arising from a perturbation of the background metric at the acoustic horizon providing a -- in principle testable -- realization of the so called black hole \textit{membrane paradigm}.
\end{abstract}

\maketitle

\section{Introduction}
Black holes (BHs) are perfect playgrounds for  quantum mechanics and general relativity given their understanding requires both~\cite{Hawking:1975vcx}.
Present quantum technologies provide accurately controllable platforms where analogue (sonic) BHs~\cite{Unruh:1980cg} can be realized~\cite{Steinhauer,Chin2017, MunozdeNova:2018fxv,Chin2019} and investigated in one-to-one correspondence with precise theoretical predictions. The latter have the advantage that can be based either on quantum fluid-dynamics, just hinging on very general conservation laws and
symmetries~\cite{Barcelo:2005fc, Mannarelli:2008jq, Carusotto, Mannarelli:2020ebs}, or on microscopic theories of interacting matter based
on accurately known microscopic Hamiltonians~\cite{zwerger2011bcs}. The  gravity analogue approach can best provide hints on those properties of systems that can be considered as universal, i.e. independent of the underlying microscopic structure, with the advantage of probing the essential ideas in combined, on-purpose designed, theoretical and experimental efforts, and the disadvantage of hardly providing clues on the microscopic-dependent details.

A number of BHs concepts have been investigated within an analogue-gravity approach, especially regarding the observation of Hawking radiation and temperature~\cite{Steinhauer,Chin2019, Unruh:1980cg,Visser:1997ux}, all supporting the universal nature of the corresponding ideas. One of the  most striking predictions of universal behavior concerns the shear viscosity-to-entropy density ratio $\eta/s$, that is conjectured  to satisfy  the so-called KSS bound $\eta /s \geq 1/4\pi$. Called after Kovtun, Son and Starinets  who first derived it within the AdS/CFT correspondence~\cite{Kovtun:2004de}, the KSS bound has been independently worked out in Rindler causal horizon in flat spacetime~\cite{Chirco:2009dc, Chirco:2010xx}. The speed of light not appearing in the bound, the KSS conjecture seems to be valid for all real fluids, relativistic and not, and can be extended to classical fluids. The highly debated question here arises, on which are the necessary conditions for the fluid to provide the  minimum $\eta/s$ value, just hinging on quantum mechanics~\cite{Rupak:2007vp, zwerger2011bcs, Adams:2012th}.

The answer to this question is relevant when bringing general relativity and quantum mechanics together at the black hole horizon, and also as a guidance to account for discrepancies between microscopic theoretical predictions~\cite{zwerger2011bcs} and experimental realizations~\cite{Cao:2010wa}. One general answer to the question comes from the KSS work itself, i.e. that the fluid must be strongly interacting with no well-defined quasi-particles~\cite{zwerger2011bcs}, something that is  found  in relativistic heavy-ion collisions close to  the deconfinement transition temperature, see for instance~\cite{Shuryak:2014zxa},
in pure gauge numerical simulations~\cite{Meyer:2007dy}, in $O(N)$ models~\cite{Romatschke:2021imm} as  well as in  hadronic models~\cite{Rais:2019chb}.
So far, hydrodynamic transport coefficients such as viscosities were mainly calculated from a microscopic theory using {Kubo-like formulas}, which involve finite temperature Green’s functions of conserved currents~\cite{ENSS,zwerger2011bcs,Chirco:2010xx}.

Here, we propose an entirely different, simpler and more transparent perspective, where the KSS bound is derived within the
viewpoint of kinetic theory~\cite{LINDQUIST1966487, stewart1969lecture, Volovik:2003ga, Mannarelli:2008jq}, enhancing purely geometric  considerations as keys to reading.
Besides widening the understanding of the essential idea and its universal nature,  our derivation may be especially useful to design analogue gravity experiments in current platforms, where the concept can be probed. In essence, we start from the idea that the low-energy excitations (phonons) of a flowing  fluid in Minkowski spacetime can be described in terms of scalar particles embedded in an effective gravitational background~\cite{Unruh:1980cg} with  emerging acoustic metric tensor determined by the fluid's properties~\cite{Bilic:1999sq,Visser:2010xv}. Since in the fluid rest frame phonons propagate at the speed of sound, a transonic flow  gives rise to the fluid analogue of a BH. Indeed, when the fluid velocity exceeds the speed of sound  a sonic black hole is produced: phonons cannot classically propagate from the supersonic region to the subsonic region.
One can then identify the acoustic horizon as the place where the fluid velocity equals the speed of sound~\cite{Unruh:1980cg,Bilic:1999sq,Visser:2010xv, Brout:1995rd,  Barcelo:2005fc, Volovik:2003ga}.
Quantum fluctuations at the acoustic horizon result in a thermal radiation of phonons~\cite{Brout:1995rd, Corley:1997pr,Saida:1999ap, Himemoto:2000zt,Unruh:2004zk,Barcelo:2005fc, Balbinot:2006ua} the sonic analogue of the Hawking radiation~\cite{Hawking:1975vcx}. Noticeably, this emission of  long-wavelength sonic vibrations at the acoustic horizon
has been  confirmed both numerically~\cite{Carusotto} and in the laboratory~\cite{Steinhauer, MunozdeNova:2018fxv} with atomic Bose-Einstein condensates.

As it was put forward by Volovik~\cite{Volovik:1999fc}, the  spontaneous phonon emission process at the horizon  produces quantum friction:  an irreversible phonon energy flux is produced  at the expense of  the bulk fluid kinetic energy. We show that such small reduction of the fluid kinetic energy can be described in terms of \textit{effective} shear and bulk viscosities defined close to the horizon. We determine the  viscosity coefficients  finding that their ratio to the phonon-entropy density equals $1/4\pi$~\cite{Kovtun:2004de}. This result is  obtained irrespective of the detailed phonon emission mechanism. To provide  a microscopic description of viscosity and to corroborate our findings, we derive the KSS bound employing the analogy between  hydrodynamics and gravity  to describe  the spontaneous emission process in terms of phonon tunneling. Furthermore, we study the shear viscosity arising from a perturbation of the background metric at the acoustic
horizon. In particular, we show that the   stretching of the acoustic horizon surface results in an entropy variation that can be described in terms of a shear viscosity coefficient. The ratio of such shear viscosity coefficient to the horizon surface entropy density  saturates the KSS bound.

The present paper is organized as follows. In Section~\ref{sec:TS} we present the theoretical setup, with emphasis on the  viscosity  and analogue gravity descriptions. In Section~\ref{sec:bulk_shear} we show that in the presence of an acoustic horizon one can define an effective shear and bulk viscosity that saturate the KSS bound. In Section~\ref{sec:tunneling} we derive the same result using a tunneling description of the phonon emission and in   Section~\ref{sec:entanglment} we obtain the shear-to-entropy ratio of the acoustic horizon by a quantum entanglement approach. We draw our conclusions in Section~\ref{sec:conclusions}. We use  natural units $\hbar = c = k_B =1$ and metric signature $(+---)$.

\section{Theoretical setup}
\label{sec:TS}
The main object of our interest is the viscous process associated to the presence of an acoustic horizon, so we briefly introduce the concepts of viscosiy coefficients and of acoustic metric.

\subsection{Shear and bulk viscosities}
The viscosity coefficients describe the non-equilibrium hydrodynamics of a fluid which has been perturbed away from  equilibrium to low-order in the velocity gradients. Consider a fluid that flows in the $x$--direction with a  small velocity gradient in the $y$--direction, $\partial_y v_x$. In the absence of viscosity, the fluid motion can be described as a  laminar flow that persists indefinitely.  The shear viscosity  tends to restore the uniform flow, thus to reduce the fluid velocity gradient $\partial_y v_x$. Accordingly, the shear viscosity coefficient, $\eta$,  can be defined as the rate of momentum transfer to the velocity gradient in the direction orthogonal to the fluid flow~\cite{huang2000statistical}. Heuristically, the shear viscosity expresses the fact that the propagation of particles across layers   with different laminar velocity results in a reduction of any fluid-flow inhomogeneity. This diffusion mechanism is the main source of viscosity  in gases and it is   simply due to the fact that particles in different laminar layers  possess  average  different velocity components along the $x$--direction. As a result, their propagation between different layers tends to reduce the average velocity difference and thus to make the fluid flow  more homogeneous. According to this mechanism,  the shear viscosity is proportional to the temperature $T$: at  high $T$, particle diffusion takes  place rapidly, see~\cite{Hosoya:1983id} for a field theory derivation.
In general, the diffusion  mechanism is very efficient when the particles' mean free path is large, meaning that the smaller are the interparticle interactions, the larger the viscosity comes out to be.  This happens because  particles with large mean free paths  propagate to large distances, thus for example back and forth between layers with very small and very large $v_x$,  eventually resulting in a quick reduction of the velocity gradient. Since the shear viscosity is inversely proportional to the interaction rate, it is in general believed that the lower limit of the shear viscosity is attained in strongly interacting systems.  This leads to the question of whether there is a fundamental lower  limit to  the shear viscosity coefficient as the strength of the interaction is increased, see for instance the discussions in~\cite{Hosoya:1983id, Danielewicz:1984ww, Kovtun:2004de, Rupak:2007vp}.

The longitudinal bulk viscosity coefficient, $\zeta$, describes
how a fluid with a small  velocity gradient $\nabla \cdot v$ restores the homogeneous flow. Similarly to the shear viscosity, it is expected that increasing gas temperatures yields to larger bulk viscosity coefficients~\cite{Hosoya:1983id}.
Both the shear and bulk viscosities describe irreversible transformations~\cite{Landau-stat12} given that they are associated with the  entropy production terms in the hydrodynamic equations:  one  can interpret the bulk viscosity coefficient as the dissipative term related to a fluid expansion or compression, while the shear viscosity as the dissipative term related to shear stresses acting on a fluid. Given the irreversible character of these processes it is widespread  practice (especially in particle physics) to use  their ratios  to the entropy density,
$\eta/s$ and $\zeta/s$,  as measures of the extent to which the fluid moves away from ideal conditions.

One important aspect is that  the bulk viscosity coefficient identically vanishes for a conformal fluid,   a fluid expansion or compression is  indeed equivalent to a fluid dilatation, which in turn is a symmetry transformation for a conformal fluid. Actually, to have vanishing bulk viscosity it is enough to have scale invariance. However, the shear viscosity of a conformal fluid is expected to be nonzero. Indeed, it has been conjectured in~\cite{Kovtun:2004de}  that the shear viscosity coefficient-to-entropy density ratio of a fluid has  the universal lower bound $1/4\pi$.

\subsection{Acoustic horizon}
\begin{figure*}[htb]
  \includegraphics[angle=0, width=0.9\columnwidth]{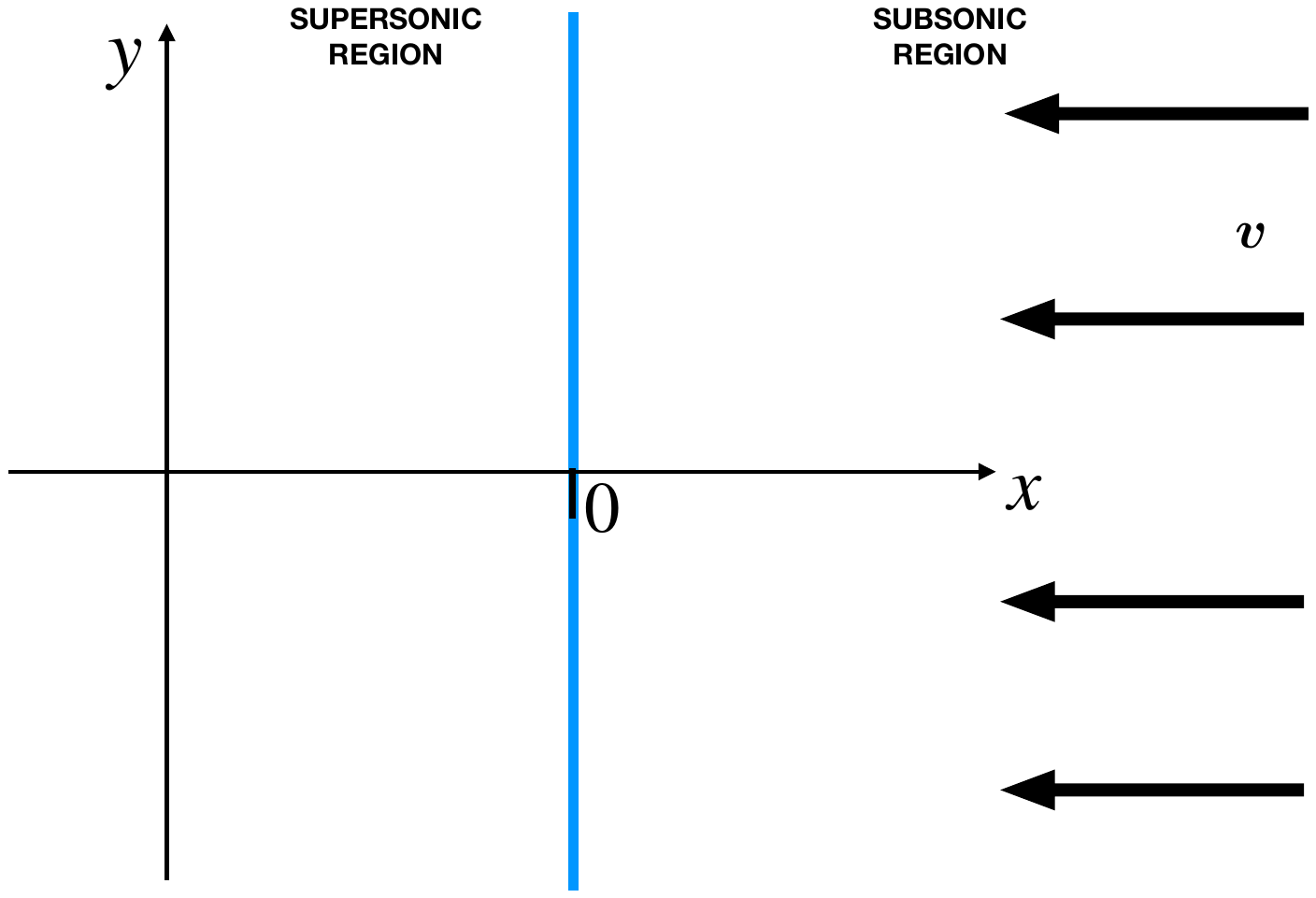}\hspace{1cm}
 \includegraphics[angle=0, width=0.9\columnwidth]{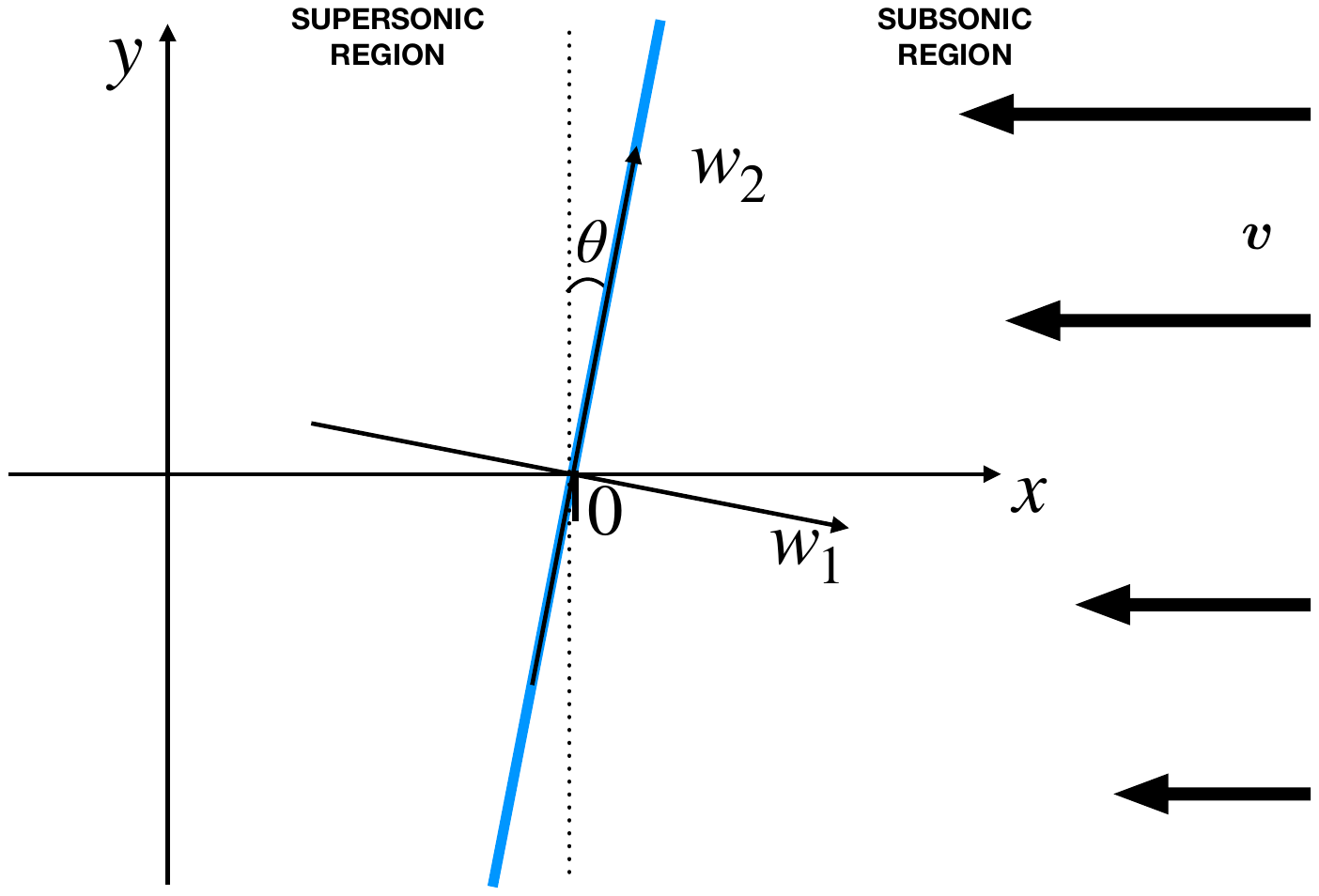}
\caption{Two-dimensional representation of the acoustic horizon determined by the fluid velocity profile given in Eq.~\eqref{eq:vx}. Left panel: the fluid velocity increases from right to left but has no shear, corresponding to  $C \neq 0$ and  $k=0$ in Eq.~\eqref{eq:vx}. The blue line  orthogonal to the $x$--axis corresponds to the acoustic horizon. Right panel:  the fluid velocity increases from right to left and from bottom to top, corresponding to $C \neq 0$ and  $k \neq 0$ in Eq.~\eqref{eq:vx}. The blue line corresponding to the horizon is now titled by the angle $\theta \simeq  -k/C$ with respect to the $y$-axis;   the coordinate axes $w_1$ and $w_2$ are  orthogonal and parallel to the tilted horizon, respectively. }
 \label{fig:horizon_tilted}
\end{figure*}

To study whether  viscous processes can arise in transonic fluids,  we consider an idealized system:  an inviscid, barotropic fluid at vanishing temperature in Minkowski spacetime.  In this setting it can be shown that  the phonon propagation can be viewed as the propagation of a scalar particle in an emergent gravitational background described by the acoustic metric~\cite{Barcelo:2005fc}
\be
\label{eq:metric}
g_{\mu\nu}= \eta_{\mu \nu} + \left(c_s^2 - 1 \right)  v_\mu  v_\nu\,,
\ee
where $\eta _{\mu\nu} = \text{diag}(1,-1,-1,-1)$ is the standard Minkowski metric, $c_s$ is the adiabatic sound speed and $v_\mu = \gamma(1, -{\bm v}) $ is the fluid four-velocity, with $\gamma$ the Lorentz factor. For definiteness, we assume that the velocity is oriented along the negative $x$-axis and its modulus is given by
\be\label{eq:vx}
v= c_s - C x + k y\,,\ee
where $C$ and $k$ are two parameters, with $k \ll C$ to ensure a negligible flow along the $y$ direction.
The last term on the r.h.s. of~\eqref{eq:vx} represents a small shear perturbation to the fluid flow.
The horizon position  is the place where the fluid velocity equals the speed of sound and it is  depicted in Fig.~\ref{fig:horizon_tilted} with a blue line. The acoustic horizon corresponds to an imaginary plane separating the supersonic and subsonic regions: no sound wave can classically propagate from the supersonic region to the subsonic one.  In the left panel of Fig.~\ref{fig:horizon_tilted} it is shown the case with $k=0$, while the right panel corresponds to a fluid  with a small shear component; in this case the horizon is tilted by the angle $\theta \simeq  -k/C$ with respect to the $y$-axis.

In analogy with standard BHs,  acoustic holes emit phonons  at the Hawking temperature
\be\label{eq:hawking}
T = \frac{1}{2 \pi} \left.\left(\frac{ c_s-|v|}{1-c_s |v| }\right)'\ \right\vert_H
\simeq \frac{C}{2 \pi} \,,
\ee
where the prime indicates the derivative with respect to the direction orthogonal to the horizon,
and the last expression is obtained in the non-relativistic limit and at  leading order in $k/C$. Expression~\eqref{eq:hawking} has been derived by a kinetic theory  approach in~\cite{Mannarelli:2020ebs} and  reproduces the standard form of $T$ in the non-relativistic limit~\cite{Barcelo:2005fc}.

We assume that the temperature of the fluid, $T_\text{fluid}$, is
negligible as compared to  $T$, that is $T_\text{fluid} \ll T$. For
simplicity, we actually set $T_\text{fluid}=0$. Thus,
for subsonic flow the fluid viscosity and entropy density vanish.
Such state of affairs changes for  transonic flow. In the presence of an acoustic horizon, the non-vanishing  Hawking temperature leads to expect nonzero viscous coefficients and  entropy density.  In this scenario, the viscosity should be related to the fact that the acoustic horizon spontaneously emits  phonons, while the entropy is the one of the phonon gas emitted by the horizon,  see~\cite{Mannarelli:2020ebs}.

In the following we limit our analysis  to a small neighbourhood $\ell$  of the horizon position, satisfying the condition
\be \label{eq:hydro_cond}
L_c \ll \ell \ll \lambda_T\, ,
\ee
with $L_c$ the theory cutoff accounting for the underlying microscopic structure invisible to our treatment; in acoustic holes realized with Bose-Einstein condensates this is e.g. of the same order of the healing length of the condensate. The length scale
\be\label{eq:lambdaT}
\lambda _T  \simeq  \frac{c_s}{2 \pi T }\,,
\ee
is  a thermal wavelength associated to the Hawking temperature.  In turn, taking into account the above expression of the Hawking temperature,   the condition $\ell \ll \lambda_T$ ensures that we consider a small neighborhood of the horizon, where  the fluid velocity is close to $c_s$.

\section{Bulk and shear viscosities in the presence of an acoustic horizon}
\label{sec:bulk_shear}
 In the absence of transonic flow, that is in the absence of the acoustic horizon, the  energy momentum tensor of the system does not contain dissipative terms and can be written as the one of a perfect fluid, that is
 \be
 T^{\mu \nu}_0 = (\rho + P ) v^\mu v^\nu  - P \eta_{\mu \nu}\,,
 \ee
where $\rho$ is the energy density and $P$ is the pressure. These two quantities are related by an equation of state that we have no need to specify.
Let us now turn to the case of transonic fluid flow. Even at zero temperature, because of the presence of  the acoustic horizon, we have to include the contribution due to the phonons  spontaneously emitted by the horizon. Thus, the total energy momentum tensor is now
 \be
 T^{\mu \nu}_\text{T} = T^{\mu \nu}_0 +{\cal T}^{\mu\nu} \,,
 \ee
 with the general expression of the phonon energy-momentum tensor  given by
\be
\label{eq:def_tmunu}
{\cal T}^{\mu\nu}   =  \int p^\mu p^\nu f_\text{ph}(x,p) d {\cal P} \,, \ee
where $f_\text{ph}$ is the phonon distribution function and $d {\cal P}$ is the appropriate covariant momentum measure~\cite{LINDQUIST1966487, stewart1969lecture}, taking into account that phonons propagate in the acoustic metric given in Eq.~\eqref{eq:metric}. 

On the other hand, we can decompose the total energy momentum tensor as the sum of two terms
\be
T^{\mu \nu}_\text{T} = T^{\mu \nu}_0 + \sigma^{\mu \nu}\,,
\ee
where $\sigma^{\mu \nu}$ is the viscous stress tensor~\cite{Landau-stat12} including all the dissipative contributions. However, since any phonon-phonon scattering process is assumed to be suppressed, the only process contributing  to $\sigma^{\mu \nu}$ is the spontaneous  phonon emission at the horizon. This is indeed an irreversible process, which transforms the kinetic energy of the fluid in phonon excitations. This is a distinctive result with respect to other existing models: essential to the setting in of a viscous horizon is just geometry, a property that we could identify by adopting the kinetic theory approach.  Assuming small perturbations of the fluid velocity, the leading terms of the  viscous stress tensor linearly depend on the velocity gradients. Moreover, at the leading order in $k/C$, we expect that  only  its
$xx$,  $yx$ and $xy$ components  are nonzero. This is due to the fact that the rotational symmetry is broken by the tilted acoustic horizon and that the horizon can  only produce a pressure in the $w_1$ direction, see the right panel of Fig.~\ref{fig:horizon_tilted}. This means, for instance, that the pressure along the $y$ direction is of order $(k/C)^2$. We shall explicitly see it  using the kinetic theory approach.
For these reasons we  assume that the leading contribution to the viscous stress tensor   is
\be\label{eq:sigmap}
\sigma_{ik} = \eta  (\delta_{iy}\delta_{kx} \partial_i v_k+\delta_{ix}\delta_{ky} \partial_k v_i) + \zeta \delta_{ix}\delta_{kx}{\bm \nabla} \cdot {\bm v}\,,
\ee
where  the velocity is given in Eq.~\eqref{eq:vx}, and  $\eta$ and $\zeta$ are the transverse and longitudinal  transport coefficients of the fluid.

We are now going to determine  $\zeta$ and  $\eta$ by equating the viscous stress tensor of the fluid to the energy-momentum tensor of the phonons,
under the working hypothesis that both $\eta$ and $\zeta$ are only due to the spontaneous emission of phonons at the horizon, thus occurring only in a thin layer around its position, see Eq.~\eqref{eq:hydro_cond}. At the end of our derivation, we will check to what extent the used approximation is valid. 

An important point to remark is that in general one cannot identify the phonon energy momentum tensor with the viscous stress tensor. For instance, in a three-dimensional homogeneous fluid the phonon thermodynamic  pressure  is isotropic, proportional to $T^4$ and it is not related to a fluid transport coefficient. The $T^4$ dependence can be viewed as arising from the fact that the  pressure is proportional to the energy density, which  can be obtained as the ratio between the average phonon energy, proportional to $T$, and the thermal wavelength cube,  proportional to $T^{-3}$.   In an infinite volume, the thermal  wavelength is indeed the only length scale of a non-interacting phonon gas. The difference, in the present case, is that we have to restrict to a small volume  close to the acoustic horizon with height $w_1$, see Fig.~\ref{fig:horizon_tilted},  constrained by~\eqref{eq:hydro_cond}. We take as basis of the volume element a patch of area $L_c^2$ determined by the microscopic cutoff of the system.  This is  the area of the smaller patch in which we can divide the horizon surface, in a way that the system can be  effectively described as  $1+1$ dimensional.  Thus,   the phonon energy density is proportional to $T$ over the elemental volume  $L_c^2 w_1$.
Moreover, $T$ is now the Hawking  temperature, meaning that it is proportional to the velocity gradient, see Eq.~\eqref{eq:hawking}. Thus, the phonon contribution to the energy momentum tensor is expected to be  directly proportional to the gradients of the velocity.   We will now quantitatively develop these ideas using the kinetic theory approach  proposed in~\cite{Mannarelli:2020ebs}.

Close to the horizon,  the phonon distribution function can be cast in the form
\be\label{eq:distribution}
f(x,p) = \frac{1}{e^{E_+/T}-1}\delta(p_{w_2})\delta(p_z) \left(\frac{2 \pi}{L_c}\right)^2\,,
\ee
 where \be\label{Eq:Ep}
E_+ = -p_x v + c_s \sqrt{p_x^2+ p_y^2 + p_z^2}\,,
\ee
 is the phonon dispersion law in the non-relativistic limit and
 $ p_{w_2} = p_y \cos\theta + p_x \sin{\theta}
$ is the momentum parallel to the horizon. The  distribution function takes into account that  phonons can only be emitted orthogonally to the horizon, see the discussion in~\cite{Mannarelli:2021olc}. Thus,    we have that
$p_z=0$, $\displaystyle p_y \simeq -\frac{k}{C} p_x$,
and  therefore the phonon dispersion law reads
$E_+ = p_x C w_1(1 + {\cal O}(k^2/C^2))$.
Upon substituting this expression in~\eqref{eq:distribution}, we can readily evaluate the leading order  components of  the phonon energy-momentum tensor. The phonon energy density
\be\label{eq:epsilon}
\epsilon_\text{ph} = \sqrt{-g} {\cal T}^0_0 \simeq \frac{T}{24 L_c^2 w_1}\,,
\ee
 takes an  expression similar to the one derived in~\cite{Mannarelli:2020ebs}, with  $L_c^2 w_1$  the small volume orthogonal to the horizon. Note that the phonon energy density has been defined in such a way that 
 \be\label{eq:Eph}
E_\text{ph} =  \int dV\epsilon_\text{ph}\,,
\ee
gives the total phonon energy. Accordingly, the phonon pressure along the $x$-direction is
\be
P_{\text{ph} \, x} = \sqrt{-g} {\cal T}^x_{x} \simeq
\frac{T}{24 L_c^2 w_1}\,,
\ee
and it is therefore equal to the phonon energy density.
The ${\cal T}^{y}_{x}$ component leads instead to a form of transverse pressure:
\be
P^{y}_{\text{ph} \,x}= \sqrt{-g} {\cal T}^y_{x} \simeq - \frac{k}C P_x \,.
\ee
Finally, the phonon pressure along the $y$-axis turns out to be:
\be
P_{\text{ph} \, y}= \sqrt{-g} {\cal T}^y_y \propto (k/C)^2\,,
\ee
and thus, as anticipated, negligible to leading order in $k/C$.  We now implement our key idea that the spontaneous phonon emission process at the horizon produces the irreversible heat increase at the expense of the bulk fluid kinetic energy described by $\eta$ and $\zeta$. This quantum friction is thus determined by the fact that momentum is transported  across the  fluid   in the direction orthogonal to the horizon by the spontaneously emitted phonons.  We thus equate the viscous stress fluid contributions, accounting for the variation of the background energy-momentum tensor, with the corresponding phonon terms: $\sigma_{xx}= P_{\text{ph} \, x}$ and $\sigma_{xy}= P^y_{\text{ph} \, x}$.
As it can be  shown using kinetic theory, see~\cite{Mannarelli:2020ebs},  the thermodynamic relation $\epsilon_\text{ph} = -P_{\text{ph} \,x} + Ts_\text{ph}$,
in the considered inhomogeneous system is valid at the leading order in $k/C$. Thus we readily have that
\be\label{eq:KSS_bound}
\frac{\zeta}{ s_\text{ph}} =\frac{\eta}{ s_\text{ph}} = \frac{1}{4 \pi}\,.
\ee
Therefore, the KSS bound is saturated for both the shear and the  bulk viscosities of the fluid~\cite{Kovtun:2004de}. A few comments are in order. These viscosity coefficients are only well defined close to the  horizon, in the region given in~\eqref{eq:hydro_cond}, where phonon self-interactions can be neglected. We have obtained this result without any reference to the microscopic emission mechanism of phonons. The above derivation is based on the assumption that the phonon emission results in  a small variation of the background energy-momentum tensor. This happens to be the case, for instance, in the experimental setup of~\cite{Steinhauer}. The shear and bulk viscosities are equal because they share a common origin. The viscosities, indeed, only arise as a consequence of the spontaneous phonon emission orthogonal to the horizon.

The latter property can be formally derived considering that 
 the energy momentum tensor in the $(x,y)$ coordinates can be obtained by a
 rotation of the one in $(w_1,w_2)$ coordinates, see the right panel of Fig.~\ref{fig:horizon_tilted}.
The phonon emission occurs along the $w_1$--direction, hence in the $(w_1,w_2)$ coordinates it reads as
\be
\label{eq:tmunu}
\sqrt{-g}{ \tilde {\cal T}}_{j}^{i}= \delta ^{i w_1}\delta_{j w_1 } P_{ \rm ph } \, ,
\ee
with  $P_\text{ph}$ the pressure of the phonon gas. The energy momentum tensor in the $(x,y)$ coordinates is obtained performing a counterclockwise rotation $R(-\theta )$ of angle $-\theta \simeq k /C$ of ${ \tilde {\cal T}}_{j}^{i}$, leading to
\be
 {\cal T}_{ \, j } ^{ i} = R (-\theta) ^i_l \,  R (-\theta ) ^m _j \,  {\tilde {\cal T}}_{m } ^{ l }  \, .
\ee
Upon using the above expression~\eqref{eq:tmunu}, the non vanishing spatial components are
\begin{align}
\sqrt{-g} {\cal T}& = \left( \begin{matrix}
\cos ^2 \theta &   -\sin \theta \cos \theta \\
-\sin \theta \cos \theta  & \sin ^2 \theta
\end{matrix} \right) \, P_\text{ph} \nonumber \\
& \simeq \left( \begin{matrix}
1 &   -\theta \\
-\theta  & 0
\end{matrix} \right) \, P_\text{ph} + {\cal O}(\theta ^2)\, , \label{eq:texpanded}
\end{align}
where in the last step we have retained only leading order terms in $\theta$.
Equating the previous expression with the viscous stress tensor $\sigma _{ij}$ in Eq.~\eqref{eq:sigmap}, and using the thermodynamic relation
$P_\text{ph} = T s_\text{ph}/2$, we obtain
\begin{align}
\zeta (\nabla \cdot {\bm v} )=&  \frac{Ts_\text{ph}}{2} + {\cal O} (\theta ^2) \, , \\
\eta \de _y v =&  - \frac{Ts_\text{ph}}{2}\theta + {\cal O} (\theta ^2)\, .
\end{align}
Replacing the expression of $v$ given in Eq.~\eqref{eq:vx} and the definition of Hawking temperature in Eq.~\eqref{eq:hawking}, we readily get that the shear and bulk viscosities are equal and satisfy Eq.\eqref{eq:KSS_bound}.

It is instructive to check whether the obtained result is consistent with the requirement that the hydrodynamic expansion is under control. As we have mentioned, the hydrodynamic description is consistent if the viscous stress tensor  linearly depends on the velocity gradient; given Eqs.~\eqref{eq:sigmap} and ~\eqref{eq:hawking} it means that $\sigma_{ij}$ should be linearly dependent on the Hawking temperature. It follows
that the present approach is  consistent if  the viscosity coefficient are independent of the temperature. In other words, since a velocity gradient is proportional  to the Hawking temperature,  we can view the viscous stress tensor as an  expansion in the Hawking temperature. Roughly speaking, this implies that
\be
\sigma \sim \zeta\,  T + {\cal O}(T^2)\,,
\ee
where we have suppressed the space indices and  taken into account that we found that $\zeta=\eta$. The present hydrodynamic approach is valid if the quadratic temperature term is suppressed with respect to the  linear term. As a matter of fact we have that
\be
\label{eq:KSS}
\zeta = \eta = \frac{s_\text{ph}}{4 \pi } = \frac{1}{48 \pi L_c^2 w_1}\,,
\ee
thus the viscosity coefficients are independent of the Hawking temperature. Moreover,  close to the horizon, they take a large value ensuring that they provide the leading contribution to the viscous stress tensor. The viscosities  decrease as the distance from the horizon $w_1$ increases signaling that the dissipative process only happens in a region close to the horizon. Note that  $w_1$ is bounded in the hydrodynamic interval~\eqref{eq:hydro_cond}, meaning that
\be\label{eq:Tdep}
\zeta = \eta \gg \frac{T}{12 L_c^2 c_s}\,,
\ee
where the value on the right hand side is obtained taking $w_1 = \lambda_T$, see Eq.~\eqref{eq:lambdaT}. In other words, as expected, at the distance  $\lambda_T$ from the horizon the proposed  hydrodynamic description breaks down. At the thermal distance $\lambda_T$ the viscosity coefficients takes  the  temperature dependence expected in case the viscosity is due to a diffusion mechanism, that is proportional to $T$. One may be tempted to  expect that the value on the right hand side of Eq.~\eqref{eq:Tdep} could be a reasonable extrapolation of the quantum viscosity to the standard hydrodynamic diffusion viscosity.  However,  one should  consider that $T$ in Eq.~\eqref{eq:Tdep} is the Hawking temperature and not the bulk thermodynamic temperature and that at the   $\lambda_T$ distance from the horizon the scattering processes between  phonons can give a sizable contribution to  viscosity.

\section{Phonon tunneling }
\label{sec:tunneling}
We have seen how the viscous coefficients rule the macroscopic behaviour of the
background fluid, i.e. its response to a perturbation of the velocity field (or of the sound velocity). We now enter the nature of this process and show that it can be interpreted as the consequence of phonons tunneling the acoustic horizon. In doing this we have to take into account that
world-lines of phonons are null geodesics of the acoustic metric, see~\eqref{eq:metric}
, and their dispersion law depends on the position with respect to the horizon,
see the definition of $E_+$ in Eq.~\eqref{Eq:Ep} and the discussion in~\cite{Mannarelli:2020ebs}.

In order to grasp the underlying physics, we use a semiclassical description of the Hawking emission process~\cite{Parikh:1999mf, Volovik:1999fc}.
The WKB tunneling amplitude is determined by the exponential of the imaginary part of the action \be \text{Im\,S} =\text{Im} \int_{x_\text{in}}^{x_\text{out}} p(x)  dx\,,\ee where $x_\text{in/out}$ are  points inside/outside the horizon and  $p(x)$ is the momentum conjugate to the $x$-coordinate; it corresponds to the phonon momentum during the emission process. As in~\cite{Parikh:1999mf}, we can  use the  Hamilton-Jacobi equation $\displaystyle \dot x ={d \omega }/{d p}$, with $\omega$  the energy of the outgoing phonon with momentum $p(x)$ to change the integration variable.
The integral can then be evaluated by using the expression of the null geodesic condition for phonons moving upstream ($x>0$) (see~\cite{Mannarelli:2020ebs})
\be\label{eq:geo1}
\dot x = \frac{c_s-v}{1-c_s v}\,.
\ee
After a change of variables, we obtain
\be\label{eq:amplitude_2}
\text{Im\,S}  = \text{Im} \int_{x_\text{in}}^{x_\text{out}} dx  \int_{0}^{\omega}   d \omega \frac{1-c_s v}{c_s-v}\, ,
\ee
where we took into account that the emitted phonon
has initially vanishing energy.
Upon  changing the order of the integrals and using the velocity profile~\eqref{eq:vx},
we obtain that the probability of quanta emission with energy $\omega$ is given by
\be
\label{eqn:thermal_prob}
P(\omega) \, \propto \, e^{-2\, \text{Im\,S} } =  e^{- \omega/{T} }\,,
\ee
which is the Boltzmann distribution with the Hawking temperature given in Eq.~\eqref{eq:hawking}.
This result reflects the fluctuation-dissipation theorem linking viscous dissipation to fluctuations of a thermal equilibrium state.
The thermal phonon emission at the horizon results in an irreversible  decrease of the kinetic energy of the fluid~\cite{Volovik:2003ga,Balbinot:2004da}, which may be included in our derivation of the emission probability. We have calculated this backreaction effect  finding that it produces  a correction to the Boltzmann distribution of the same form obtained in~\cite{Parikh:1999mf}, with the black hole mass replaced by the fluid mass times the speed of sound squared. Notice that introducing the backreaction of the acoustic horizon would not alter the main idea we are focusing on. For this reason we leave that discussion  to a future publication.

We instead proceed now to describe the dissipative processes associated to the spontaneous phonon emission. In particular, we  determine the viscosity coefficients by equating the kinetic energy loss of the fluid with the phonon energy gain, corresponding to the heat emitted by the horizon. The total energy of the phonons, see Eq.~\eqref{eq:Eph}, considering  a small  volume close to  the horizon having as basis a patch of area $L_c^2$ is
\be
E_\text{ph} = \int \epsilon_\text{ph} d^3 x \simeq L_c^2 \int_{L_c}^{w_1}  \epsilon_\text{ph} d w_1 = \frac{T}{24} \log\left(\frac{w_1}{L_c}\right)\,,
\ee
where the phonon energy density is given in~\eqref{eq:epsilon}.
Thus, the emission power of   the horizon is
\be\label{eq:geodeps_z1}
 \dot E_\text{ph} =  \frac{T}{24} \frac{\dot w_1}{w_1} =\epsilon_\text{ph} L_c^2  \dot w_1 \,,
\ee
where
\be
\label{eq:zperpdot}
\dot w_1  = w_1 \frac{C^2+k^2}{C}\,,
\ee
which is obtained  after taking  the non-relativistic limit of~\eqref{eq:geo1} and considering  the leading order in
$k/C$.

The  rate of kinetic energy loss of the fluid due to dissipative processes is instead
\begin{align}\label{eq:}
\dot E_\text{fluid} &= - \int d^3 x \frac{\sigma_{ik} }{2} \left( \partial_i v_k +\partial_k v_i \right)\nonumber \\ & \simeq - L_c^2 w_1 \frac{\sigma_{ik} }{2} \left( \partial_i v_k +\partial_k v_i \right) \,,
\end{align}
where the viscous stress tensor is given in Eq.~\eqref{eq:sigmap} and in the last expression we only considered  the relevant  thin layer close to the horizon.
Upon equating the kinetic energy loss of the fluid with the energy gain of the phonon gas we obtain once again Eq.~\eqref{eq:KSS_bound}.
\section{$\eta/s$ of the horizon and its relation to quantum entanglement}
\label{sec:entanglment}
In the previous sections  we have derived the hydrodynamic viscous coefficients close to the acoustic horizon and their ratio with the entropy density of the phonon gas. First, we have derived Eqs.~\eqref{eq:KSS_bound} and~\eqref{eq:KSS} by a general kinetic theory approach. Then, in Section~\ref{sec:tunneling}, we have interpreted the spontaneous emission of phonons at the acoustic horizon as a tunneling process finding the same results. In both approaches we have thus derived the viscosity coefficients of the fluid determined by the spontaneous phonon emission at the horizon.

In the following, we focus on the ratio between the shear viscosity coefficient and the entropy {\it surface density} of the horizon {\it itself} and its interpretation as entanglement entropy of the acoustic horizon. 
 Note that the phonon entropy and the horizon entropy  are two distinct quantities. The former is the thermodynamic entropy of the (phonon) gas, the latter is instead associated to the horizon area. However, as we have previously seen, a variation of the acoustic horizon area determines a change of the phonon entropy density since we assume the background fluid to carry, or produce,  zero or negligible entropy. This means that the variations of the two entropies are connected.
The shear perturbation of the background fluid (hence of effective spacetime) turns into a deformation of the horizon producing a variation of the cross-sectional area element of the horizon projected onto the holographic screen ($yz$ plane), which is orthogonal to the plane shown in  Fig.~\ref{fig:horizon_tilted}.
We notice that similarly to BHs, the presence of the acoustic horizon produces a spatial partition of the vacuum state giving rise to an entanglement entropy~\cite{Finazzi:2013sqa}. In our effective framework, this entropy must be normalized by the cutoff scale $L_c$ hence it scales with the horizon area: $\displaystyle S_H = \kappa {A_H}/{L_c^{2}}$.
For $\kappa = 1/4$ and identifying $L_c$ with the Planck length, this expression coincides with the Bekenstein-Hawking entropy of BHs~\cite{Bekenstein:1973ur}, which indeed has been interpreted as an entanglement entropy~\cite{Sorkin:1984kjy,Srednicki:1993im} (see also~\cite{Jacobson:2015hqa} and ref.s therein). Whether the entanglement  entropy and the (analogue) BH entropy  are actually the same is matter of debate. For our purposes, it suffices to refer to the entanglement entropy, which is in general proportional to the area separating two sub-systems~\cite{Sorkin:1984kjy}.

To produce an horizon tilt perturbation we consider  the velocity profile of Eq.~\eqref{eq:vx} with  a time-dependent shear velocity, that is ${\dot k} \neq 0$. This can be thought as  originated by an external perturbation of the fluid velocity or of the sound speed. Given the expression of the acoustic metric in Eq.~\eqref{eq:metric} this corresponds to a metric perturbation and  effectively  gives rise to a time dependent entropy surface density as seen  on a screen perpendicular to the fluid flow.
Using $y =x_0  - {kx}/{C}$, see the right panel of Fig.~\ref{fig:horizon_tilted}, it readily follows that the area change is given by
\be\label{eq:dAdt}
\frac{d A_H}{d t} = \frac{k \dot k}{C^2}\,d y dz\,,
\ee
and the corresponding entropy change is thus $\displaystyle \dot S_H = \kappa \dot A_H/L_c^{2}$.

The deformation of the horizon has its counterpart on the variation of the phonon world-lines, which should always remain orthogonal to the horizon.
Thus, in analogy with the membrane paradigm for BHs~\cite{membrane}, and under the condition expressed by~\eqref{eq:hydro_cond},
we identify these phonons as a fluid endowed with a shear
in a ``effectively'' 2+1 dimensional thin layer about the horizon.
In this picture, the entropy {change} of the horizon is transferred to the emitted phonons, giving rise to a shear viscosity of the fluid.
Following~\cite{Chirco:2010xx}, we may call it the {\it entanglement viscosity} of the horizon. Now, the phonons are emitted along $w_1$, see the right panel of Fig.~\ref{fig:horizon_tilted}. Their velocity was given  in~\eqref{eq:zperpdot}, thus the
components of the phonon fluid velocity are
\be
V_x = {\dot w_1} \frac{C}{\sqrt{C^2 + k^2}}\quad V_y = {\dot w_1} \frac{k}{\sqrt{C^2 + k^2}}\quad V_z = 0\, .
\ee
Finally, the shear of the phonon flow can be readily computed to be
$\displaystyle \left( \partial_i V_k+\partial_k V_i\right)^2  \simeq  2 \frac{k \dot k}{C}$.
Hence, using  the constitutive equations
\be
\frac{ds_H}{dt} = \frac{2\eta_H}{T} \left(\partial_i V_k+\partial_k V_i\right)^2 ,
\label{eq:dissipation_rate}
\ee
we obtain $\displaystyle \eta_H = \frac{\kappa}{L_c^2} \frac{1}{4\pi}$,
which has dimension of $1/{\rm length}^{2}$.
Eq.~\eqref{eq:dissipation_rate} gives the dissipation rate of the background (gravitational) shear which is transferred to the internal degrees of freedom -- associated to the fundamental scale $L_c$ -- of the horizon.

Since the horizon entropy density is  $s_H= \kappa/ L_c^{2}$, we readily obtain \be
\frac{\eta_H}{s_H} = \frac{1}{4\pi}\,,
\ee
which is the KSS ratio.
This result -- which is obtained here for the first time in the context of analogue gravity -- strengthens the hypothesis that the KSS ratio could be a fundamental holographic property of space-time (rather than only of the AdS/BH  solutions).

\section{Conclusions}
\label{sec:conclusions}
 We have shown that the conjectured universal Kovtun-Son-Starinets~\cite{Kovtun:2004de} bound of the shear viscosity coefficient to entropy density ratio  is saturated  in the region close to the acoustic horizon of a  transonic flow. The same happens for  the longitudinal bulk viscosity coefficient. We have first obtained these results by  a  kinetic theory approach where phonons spontaneously emitted at the horizon irreversibly produce  heat. This derivation is quite general, since it is not based on a microscopic emission process. Then, we have obtained the same results considering  a microscopic description of viscosity where the emission process is due to phonons tunneling the acoustic horizon. Finally, we have as well obtained the limiting value of   $\eta/s$  considering the variation of the surface area of the horizon induced by a time dependent perturbation.  The latter derivation bears some resemblance with the one proposed for a Rindler causal horizon in flat spacetime~\cite{Chirco:2010xx}.  Since in all considered cases there is no known holographic (gauge/gravity) duality, a natural explanation for this viscosity is in the peculiar properties of quantum entanglement {rather than}
in the existence of a gravity dual~\cite{Kovtun:2004de}.

Our approach makes clear that the viscosity arises from the thermal (Hawking) phonon emission  associated to the horizon entanglement entropy, shedding light on its geometrical origin and on how viscosity acts as an effective coupling between background (metric) perturbations and the acoustic horizon. This perspective helps conceptualizing proper analogue experiments, where the saturation of the KSS bound can be explored in  controllable manners. For example, the simplest situation can be envisioned to be that of a Bose-Einstein condensate in effectively 2D circular geometry~\cite{Chin2019}, by  impressing a fluid (or, equivalently, a sound) velocity profile according to~\eqref{eq:vx} and Fig.~\ref{fig:horizon_tilted}.  In any case, a precise experimental test of our findings requires  local measurements of entropy density and viscosity. The former may be achieved by employing  recent local-thermometry techniques based on radio-frequency spectroscopy via the fluctuation-dissipation theorem, as first proposed in~\cite{PhysRevLett.125.113601}. The latter may be obtained  by the  method earlier proposed in~\cite{PhysRevLett.115.020401}, where the local shear viscosity is obtained by an appropriate inversion of the cloud-averaged viscosity. A combination of local-thermometry techniques and space-dependent two-fluid equations from microscopic time-dependent density functional theory~\cite{CHIOFALO1998188} could also be envisioned.
In such experiments, our derivation can help disentangling interaction-driven and geometric originated effects.
Since there exist peculiar BH realizations that violate the KSS bound~\cite{Kats:2007mq, Brigante:2007nu, Feng:2015oea, Brito:2019ose, Bravo-Gaete:2020lzs}, it would be interesting to check whether one of the proposed geometries could be emulated by a gravitational analogue system. \\

\noindent \textit{Acknowledgments -} We thank M.~Bravo, L.~Lepori, S.~Liberati,   C.~Manuel, A.~Trombettoni and W.~Unrhu for useful comments and suggestions. \\M.L.C. acknowledges support
from the National Centre on HPC, Big Data and Quantum Computing—SPOKE 10 (Quantum
Computing) and received funding from the European Union Next-GenerationEU—National Recovery
and Resilience Plan (NRRP)—MISSION 4 COMPONENT 2, INVESTMENT N. 1.4—CUP
N. I53C22000690001. This research has received funding from the European Union’s Digital Europe
Programme DIGIQ under grant agreement no. 101084035. M.L.C. also acknowledges support from
the project PRA\_2022\_2023\_98 “IMAGINATION”, from the MIT-UNIPI program, and in part by
grants NSF PHY-1748958 and PHY-2309135 to the Kavli Institute for Theoretical Physics (KITP).


\providecommand{\newblock}{}

\end{document}